\documentclass[final,authoryear,5p,times,twocolumn]{elsarticle}
\pdfoutput=1
\usepackage[latin9]{inputenc}
\usepackage{url}
\usepackage{amsmath}
\usepackage{amssymb}
\usepackage{graphicx}
\usepackage{lineno}
\usepackage{hyperref}

\journal{Planetary and Space Science}
\newcommand*\aap{Astronom.~Astrophys.}
\newcommand*\aj{Astron.~J.}

\newcommand*\apjl{Astrophys.~J.~Lett.}
\newcommand*\icarus{Icarus}

\newcommand*\nat{Nature}
\newcommand*\sci{Science}

\newcommand*\jgrp{J.~Geophys.~Res.~(Plan.)}

\begin{document}

\begin{frontmatter}

\title{The first confirmation of V-type asteroids among the Mars crosser
population\tnoteref{based}}

\tnotetext[based]{Based on observations obtained at the Southern Astrophysical Research
Telescope (SOAR), which is a joint project of the Ministry of Science,
Technology and Innovation of Brazil, the U.S. National Optical Astronomy
Observatory, the University of North Carolina at Chapel Hill, and
the Michigan State University.}

\author[add1]{A. O. Ribeiro}
\ead{anderson@on.br}

\author[add1]{F. Roig\corref{cor}}
\ead{froig@on.br}
\cortext[cor]{Corresponding author}

\author[add2,add3]{M. Cañada-Assandri}

\author[add1]{J. M. F. Carvano}

\author[add1]{F. L. Jasmin}

\author[add1,add4]{A. Álvarez-Candal}

\author[add2,add3]{R. Gil-Hutton}

\address[add1]{Observatório Nacional, Rua General José Cristino 77, Rio de Janeiro,
20921-400, Brazil}

\address[add2]{Universidad Nacional de San Juan, Av. España 1512 sur, San Juan,
J5402DSP, Argentina}

\address[add3]{Complejo Astronómico El Leoncito (CONICET), Av. España 1512 sur,
San Juan, J5402DSP, Argentina}

\address[add4]{Instituto de Astrofísica de Andalucía - CSIC, Glorieta de la Astronomía
s/n, E18008, Granada, Spain}

\begin{abstract}
The Mars crossing region constitutes a path to deliver asteroids from
the Inner Main Belt to the Earth crossing space. While both the Inner
Main Belt and the population of Earth crossing asteroids contains
a significant fraction of asteroids belonging to the V taxonomic class,
only two of such V-type asteroids has been detected in the
Mars crossing region up to now. In this work, we systematically searched
for asteroids belonging to the V class among the populations of Mars
crossing asteroids, in order to support alternative paths to the delivery
of this bodies into the Earth crossing region. We selected 18 candidate
V-type asteroids in the Mars crossing region using observations contained
in the Sloan Digital Sky Survey Moving Objects Catalog. Then, we observed
4 of these candidates to take their visible spectra using the Southern
Astrophysical Research Telescope (SOAR). We also performed the numerical
simulation of the orbital evolution of the observed asteroids. We
confirmed that 3 of the observed asteroids belong to the V class,
and one of these may follow a path that drives it to an Earth collision
in some tens of million years.
\end{abstract}


\end{frontmatter}


\section{Introduction}

Mars crosser (MC) asteroids are distributed in the region with $1.3<q<1.666$
AU and $a<3.2$ AU, where $q$ is the perihelion distance and $a$
is the orbital semi-major axis. Together with the Earth-crosser (EC)
asteroids ($q<1.3$ AU and $Q>0.983$ AU, where $Q$ is the aphelion
distance), they have been recognized as the potential sources of the
meteorites recovered on the Earth. The origin of these two populations
relies on the dynamical interaction between the web of mean motion
and secular resonances in the asteroid Main Belt and the thermal re-emission
forces acting on the surface of the small asteroids, the so called
Yarkovsky and YORP effects. This interaction produces a continuous
loss of asteroids especially from the Inner Main Belt (IMB; $q>1.666$
AU and $2.0<a<2.5$ AU) to the planet crossing region. Therefore,
it should be expected that the IMB, MC and EC populations show similarities
in their taxonomical distribution. In this work, we address the occurrence
of asteroids belonging to the V taxonomic class among the population
of Mars crossers.

Taxonomy allows to classify the asteroids according to observations
related to their surface properties, like colors, spectra and albedos.
Although in general there is no direct relation between taxonomic
class and mineralogy, taxonomic classification imposes some constraints
on the possible mineralogy of the body. Among the recognized taxonomic
classes (e.g. \citealp{2002Icar..158..146B}), the V class is particularly
interesting. Its spectrum shows a steep slope downwards of $\sim0.8$
$\mu$m and a deep absorption band long-wards of $\sim0.8$ $\mu$m
and centered at $\sim1.0$ $\mu$m. This band is associated to a mineralogy
typical of basalt (e.g. \citealp{2001M&PS...36..761B}). The V class
is mostly found among the members of a dynamical family in the IMB,
the Vesta family \citep{1993Sci...260..186B}, which originated from
a collision that excavated a huge crater on the basaltic surface of
asteroid (4) Vesta \citep{1970Sci...168.1445M,1997M&PS...32..965A}.
Recent results from the Dawn mission, confirm the origin of
the diogenite meteorites from Vesta's Rheasilvia crater 
(\citealp{2011Icar..212..175R}, \citeyear{2012Sci...336..700R}, 
\citeyear{2013Icar..226.1103R}; \citealp{2013JGRE..118..335M}), 
and help to interpret the differences between HED meteorites, 
vestoids and Vesta in terms of grain size and collisional implantation 
of material on Vesta's surface (\citealp{2013JGRE..118.1991B}).
The spectral peculiarities and the spatial concentration of the V-type
asteroids make them especially useful as tracers of the dynamical
paths that may transport asteroids from the IMB to the planet crosser
region.

It has been proposed (see \citealp{2002aste.conf..409M} and references
therein) that the EC and MC asteroids would be the fragments of large
bodies of the Main Belt that after a collision were injected into
the J3:1 mean motion resonance or the $\nu_{6}$ secular resonance.
The chaotic evolution inside theses resonances excited the fragments'
orbital eccentricities driving them to cross the orbits of the terrestrial
planets. The flux of asteroids falling into the J3:1, $\nu_{6}$ and
other major resonances might be continuously resupplied by the mobility
in semi-major axis caused by the Yarkovsky effect (\citealp{1998Icar..132..378F};
\citealp{2000Natur.407..606V}). \citet{1998Sci...281.2022M} and
\citet{2000Icar..145..332M} also showed that a significant fraction
of the IMB asteroids may become MCs over 100 My time scales through
weaker resonances other than the J3:1 and the $\nu_{6}$.

An analysis of the taxonomical distribution in the planet crossing
region has been done by \citet{2004Icar..170..259B} using spectroscopic
data of 254 EC and MC asteroids taken from the SMASS survey. They
found a significant correlation between the distribution of taxonomic
classes in the Main Belt and the EC and MC populations, especially
for the classes belonging to the S and Q complexes. Their analysis
also showed an albedo dependent distribution with heliocentric distance
for the classes belonging to the X complex. These results are in line
with the proposed origin of the planet crossing asteroids from Main
Belt sources, possibly with a minor contribution from other sources
like the Jupiter family comets. \citet{2010A&A...517A..23D} also
used a spectroscopic survey of other 74 EC and MC asteroids to compare
their mineralogy with that of the Main Belt asteroids and of the ordinary
chondrite meteorites. They found that EC and MC asteroids appear to
have less reddish surfaces compared to the Main Belt bodies, and that
the main source for these asteroids would be the IMB. 
\citet{2010Icar..208..773M} analyzed the near infra red (NIR) 
spectra of 39 IMB asteroids and compare them to the HED sample of 
meteorites. They found that the asteroids' mineralogy is not totally 
compatible with that of the HEDs, and they attribute this to the 
effect of space weathering, among other possibilities. In their 
sample, they include two MC asteroids, (1468) Zomba and (33881) 2000 JK66, 
compatible with a basaltic mineralogy, although the authors did not 
address these objects as Mars crossers. More recently, 
\citet{2013Icar..225..131S} analyzed the visible and
near infra-red spectra of 14 EC and MC asteroids and found that their
composition is consistent with either ordinary chondrites or basaltic
achondrites meteorites.

It must be noted, however, that all the above spectroscopic surveys
failed, in general, to identify asteroids of the V class
among the MC population, whereas about 10 \% of the IMB population
and 4 \% of the EC population with currently known taxonomy belong
to this class. \citet{2004Icar..170..259B} argued that the lack of
V-type asteroids among the MC asteroids is in line with the idea that
these bodies would be directly injected into the J3:1 and $\nu_{6}$
resonances from low eccentricity and low inclination orbits, compatible
with the Vesta family locus. Once inside these resonances, they would
rapidly evolve to EC orbits without having any dynamical interaction
with Mars. Notwithstanding, long term orbital simulations by 
\citet{2008Icar..193...85N} and \citet{2008Icar..194..125R} 
indicate that about 8 \% of the Vesta family members could become 
MCs over 2 Gy of evolution. Since the Vesta family currently has more 
than 10,000 members and it is older than at least 1.2 Gy 
\citep{2005A&A...441..819C}, we should expect to find significant 
traces of V-type asteroids in the MC region. 

In this work, we use the observations provided by the 4th. release
of the Sloan Digital Sky Survey Moving Objects Catalog (SDSS-MOC4,
\citealp{2001AJ....122.2749I}; \citealp{2002AJ....124.1776J}) to
identify MC asteroids with surface colors compatible to those of the
V class. Our searching method is summarized in Sect. \ref{sec2-1}.
In Sect. \ref{sec2-2}, we present the spectroscopic observations
of four of these asteroids, made with the SOAR telescope, in order
to confirm if they are actual V-type asteroids. An analysis of the
dynamical evolution of these bodies is given in Sect. \ref{sec3}.
Finally, Sect. \ref{sec4} is devoted to the conclusions. 

\section{V-type asteroids in the Mars Crosser region}

\begin{table}
\caption{Number of asteroids in the three dynamical regions analyzed: $N_{Ast}$
is the number of known asteroids in the ASTORB database, $N_{Tax}$
is the number of asteroids with taxonomic classification in the SDSS-MOC4
\citep{2010A&A...510A..43C}, $N_{V}$ is the number of asteroids
classified as V-type by these authors.}
\label{table0}
\bigskip{}
\centering{}%
\begin{tabular}{lccc}
\hline 
 & EC & MC & IMB\tabularnewline
\hline 
$N_{Ast}$ & 8,268 & 9,912 & 169,274\tabularnewline
$N_{Tax}$ & 91 & 533 & 19,946\tabularnewline
$N_{V}$ & 3 & 10 & 2,623\tabularnewline
\hline 
\end{tabular}
\end{table}

The shortage of V type asteroids among the MC population,
when compared to the EC, might be related to an observational bias
due to the small fraction of asteroids with observed spectra in the
two populations (1 \% among the MCs and 4 \% among the ECs). Moreover,
different spectroscopic surveys could be affected by selection effects
and completeness effects that are difficult to quantify. Nevertheless,
we can apply a simple argument to show that a significant amount of
V-type asteroids should be spectroscopically detected in the MC region. 

This argument is based on the taxonomic classification developed by
\citet{2010A&A...510A..43C} for the observations contained in the
SDSS-MOC4. Let $N_{Ast}$ be the number of known asteroids listed,
for example, in the ASTORB database
(\url{ftp://ftp.lowell.edu/pub/elgb/astorb.html })%
\footnote{The ASTORB file is produced at the Lowell Observatory, 
by Edward Bowell.}. Let $N_{Tax}$ be the number of asteroids in the 
SDSS-MOC4 that have got a taxonomic class in \citet{2010A&A...510A..43C}; 
and let $N_{V}$ be the number of these SDSS-MOC4 asteroids specifically 
classified as V-type. Table \ref{table0} summarizes these numbers for each 
population of interest. At first glance, it appears that there are more V-type
asteroids detected in the MC region than in the EC region. However,
we must bear in mind that the detection efficiency of the SDSS, which
is given approximately by the fraction $N_{Tax}/N_{Ast}$, is different
in each population, and in particular it is larger in the IMB. Therefore,
if we assume that the fraction $N_{V}/N_{Tax}$ is representative
of the actual fraction of V-type asteroids in each population, the
actual number $\bar{N_{V}}$ of V-type asteroids that the SDSS-MOC4
could have detected in either the MC or EC region is: 
\[
\bar{N}_{V}=\frac{N_{V}}{N_{Tax}}N_{Ast}\frac{N_{Tax}^{\mathrm{IMB}}}{N_{Ast}^{\mathrm{IMB}}}
\]
This gives $\bar{N}_{V}=22$ in the MC population and $\bar{N}_{V}=32$
in the EC region, and although there would be less V-type asteroids
in the MC than the EC region, their amount should be significant enough
to allow their spectroscopic detection provided that a systematic
search for these bodies is carried out. 

\subsection{Selecting candidate V-type observations from the SDSS-MOC4}
\label{sec2-1}

\begin{table*}
\caption{The targets selected for spectroscopic observation. Asteroids marked
with an ({*}) were observed spectroscopically in this work. Columns
give osculating perihelion distance ($q$), semi-major axis ($a$),
eccentricity ($e)$, and inclination ($I$) correspond to JD $2456000.5$
(2012-03-14). Table also gives the absolute magnitude $H$, the number
of SDSS-MOC4 observations considered $N_{SDSS}$, the average values
of the first and second principal components $PC_{1}$ and $PC_{2}$,
and the average Sloan $i-z$ color. The last column identifies the
taxonomic classification according to \citet{2010A&A...510A..43C};
the +NIR flag indicates that the object has been also observed
in the near infra red by \citet{2010Icar..208..773M}.}
\label{table1}
\bigskip{}
\centering{}%
\begin{tabular}{lcccccccccc}
\hline 
{\scriptsize{Asteroid name}} & {\scriptsize{$q$ {[}AU{]}}} & {\scriptsize{$a$ {[}AU{]}}} & {\scriptsize{$e$}} & {\scriptsize{$I$ {[}deg{]}}} & {\scriptsize{$H$}} & {\scriptsize{$N_{SDSS}$}} & {\scriptsize{$PC_{1}$}} & {\scriptsize{$PC_{2}$}} & {\scriptsize{$i-z$}} & {\scriptsize{Taxonomy}}\tabularnewline
\hline 
{\scriptsize{(1468) Zomba ({*})}} & {\scriptsize{1.601}} & {\scriptsize{$2.1958$}} & {\scriptsize{$0.2707$}} & {\scriptsize{$9.9434$}} & {\scriptsize{$13.6$}} & {\scriptsize{1}} & {\scriptsize{$-0.1218$}} & {\scriptsize{$-0.2182$}} & {\scriptsize{$-0.21$}} & {\scriptsize{Q (+NIR)}}\tabularnewline
{\scriptsize{(16147) Jeanli}} & {\scriptsize{1.777}} & {\scriptsize{$2.2801$}} & {\scriptsize{$0.2206$}} & {\scriptsize{$5.9068$}} & {\scriptsize{$14.2$}} & {\scriptsize{8}} & {\scriptsize{$-0.1968$}} & {\scriptsize{$-0.3125$}} & {\scriptsize{$-0.43$}} & {\scriptsize{V}}\tabularnewline
{\scriptsize{(28985) 2001 MP5}} & {\scriptsize{1.742}} & {\scriptsize{$2.2493$}} & {\scriptsize{$0.2254$}} & {\scriptsize{$4.1578$}} & {\scriptsize{$14.8$}} & {\scriptsize{1}} & {\scriptsize{$-0.1611$}} & {\scriptsize{$-0.2593$}} & {\scriptsize{$-0.41$}} & {\scriptsize{V}}\tabularnewline
{\scriptsize{(31415) 1999 AK23 ({*})}} & {\scriptsize{1.662}} & {\scriptsize{$2.2751$}} & {\scriptsize{$0.2695$}} & {\scriptsize{$6.8439$}} & {\scriptsize{$14.4$}} & {\scriptsize{1}} & {\scriptsize{$-0.1775$}} & {\scriptsize{$-0.3109$}} & {\scriptsize{$-0.40$}} & {\scriptsize{V}}\tabularnewline
{\scriptsize{(32008) 2000 HM53 ({*})}} & {\scriptsize{1.769}} & {\scriptsize{$2.1918$}} & {\scriptsize{$0.1929$}} & {\scriptsize{$6.3020$}} & {\scriptsize{$14.1$}} & {\scriptsize{1}} & {\scriptsize{$-0.2349$}} & {\scriptsize{$-0.2619$}} & {\scriptsize{$-0.42$}} & {\scriptsize{V}}\tabularnewline
{\scriptsize{(33881) 2000 JK66}} & {\scriptsize{1.567}} & {\scriptsize{$2.2124$}} & {\scriptsize{$0.2916$}} & {\scriptsize{$11.1721$}} & {\scriptsize{$14.4$}} & {\scriptsize{1}} & {\scriptsize{$-0.1167$}} & {\scriptsize{$-0.2277$}} & {\scriptsize{$-0.38$}} & {\scriptsize{V (+NIR)}}\tabularnewline
{\scriptsize{(44798) 1999 TL191}} & {\scriptsize{1.752}} & {\scriptsize{$2.2809$}} & {\scriptsize{$0.2318$}} & {\scriptsize{$7.4329$}} & {\scriptsize{$15.0$}} & {\scriptsize{2}} & {\scriptsize{$-0.1933$}} & {\scriptsize{$-0.2817$}} & {\scriptsize{$-0.40$}} & {\scriptsize{V}}\tabularnewline
{\scriptsize{(60669) 2000 GE4}} & {\scriptsize{1.765}} & {\scriptsize{$2.2066$}} & {\scriptsize{$0.1999$}} & {\scriptsize{$7.6918$}} & {\scriptsize{$15.2$}} & {\scriptsize{2}} & {\scriptsize{$-0.1775$}} & {\scriptsize{$-0.2322$}} & {\scriptsize{$-0.34$}} & {\scriptsize{V}}\tabularnewline
{\scriptsize{(67621) 2000 SY175}} & {\scriptsize{1.760}} & {\scriptsize{$2.1780$}} & {\scriptsize{$0.1919$}} & {\scriptsize{$5.4648$}} & {\scriptsize{$15.8$}} & {\scriptsize{2}} & {\scriptsize{$-0.1784$}} & {\scriptsize{$-0.2992$}} & {\scriptsize{$-0.46$}} & {\scriptsize{V}}\tabularnewline
{\scriptsize{(89137) 2001 UD17}} & {\scriptsize{1.649}} & {\scriptsize{$2.1997$}} & {\scriptsize{$0.2501$}} & {\scriptsize{$7.4623$}} & {\scriptsize{$16.6$}} & {\scriptsize{1}} & {\scriptsize{$-0.0382$}} & {\scriptsize{$-0.2743$}} & {\scriptsize{$-0.29$}} & {\scriptsize{-}}\tabularnewline
{\scriptsize{(100316) 1995 MM2}} & {\scriptsize{1.588}} & {\scriptsize{$2.2117$}} & {\scriptsize{$0.2499$}} & {\scriptsize{$4.8471$}} & {\scriptsize{$16.0$}} & {\scriptsize{1}} & {\scriptsize{$-0.1278$}} & {\scriptsize{$-0.1744$}} & {\scriptsize{$-0.13$}} & {\scriptsize{Q}}\tabularnewline
{\scriptsize{(102803) 1999 VA169}} & {\scriptsize{1.633}} & {\scriptsize{$2.1359$}} & {\scriptsize{$0.2353$}} & {\scriptsize{$4.8482$}} & {\scriptsize{$16.7$}} & {\scriptsize{3}} & {\scriptsize{$-0.2055$}} & {\scriptsize{$-0.1936$}} & {\scriptsize{$-0.19$}} & {\scriptsize{Q}}\tabularnewline
{\scriptsize{(130988) 2000 WT141 ({*})}} & {\scriptsize{1.738}} & {\scriptsize{$2.4606$}} & {\scriptsize{$0.2938$}} & {\scriptsize{$10.2803$}} & {\scriptsize{$15.3$}} & {\scriptsize{1}} & {\scriptsize{$-0.2044$}} & {\scriptsize{$-0.2280$}} & {\scriptsize{$-0.35$}} & {\scriptsize{V}}\tabularnewline
{\scriptsize{(265117) 2003 UQ26}} & {\scriptsize{1.773}} & {\scriptsize{$2.1979$}} & {\scriptsize{$0.1932$}} & {\scriptsize{$7.6565$}} & {\scriptsize{$16.8$}} & {\scriptsize{1}} & {\scriptsize{$-0.0925$}} & {\scriptsize{$-0.1971$}} & {\scriptsize{$-0.35$}} & {\scriptsize{V}}\tabularnewline
{\scriptsize{(276400) 2002 XS45}} & {\scriptsize{1.579}} & {\scriptsize{$2.1937$}} & {\scriptsize{$0.2799$}} & {\scriptsize{$5.6344$}} & {\scriptsize{$16.2$}} & {\scriptsize{1}} & {\scriptsize{$-0.2314$}} & {\scriptsize{$-0.1983$}} & {\scriptsize{$-0.19$}} & {\scriptsize{SQ}}\tabularnewline
{\scriptsize{(285894) 2001 QV25}} & {\scriptsize{1.702}} & {\scriptsize{$2.2241$}} & {\scriptsize{$0.2346$}} & {\scriptsize{$8.0129$}} & {\scriptsize{$16.8$}} & {\scriptsize{1}} & {\scriptsize{$-0.0498$}} & {\scriptsize{$-0.2270$}} & {\scriptsize{$-0.28$}} & {\scriptsize{-}}\tabularnewline
{\scriptsize{$\qquad\qquad$1999 SO15}} & {\scriptsize{1.571}} & {\scriptsize{$2.2316$}} & {\scriptsize{$0.2960$}} & {\scriptsize{$8.5455$}} & {\scriptsize{$17.4$}} & {\scriptsize{1}} & {\scriptsize{$-0.1689$}} & {\scriptsize{$-0.2404$}} & {\scriptsize{$-0.33$}} & {\scriptsize{V}}\tabularnewline
{\scriptsize{$\qquad\qquad$2005 SF57}} & {\scriptsize{1.724}} & {\scriptsize{$2.2484$}} & {\scriptsize{$0.2334$}} & {\scriptsize{$3.8331$}} & {\scriptsize{$18.1$}} & {\scriptsize{1}} & {\scriptsize{$-0.2070$}} & {\scriptsize{$-0.2303$}} & {\scriptsize{$-0.46$}} & {\scriptsize{V}}\tabularnewline
\hline 
\end{tabular}
\end{table*}

In this work, we use the five band photometric observations of the
SDSS-MOC4 survey to produce a list of potentially interesting targets
for further spectroscopic observations among the Mars crossing population.
The SDSS-MOC4 provides calibrated magnitudes of 220,101 moving objects
in the bands $u,g,r,i,z$, with their corresponding errors. These
observations are effectively linked to 104,409 unique asteroids. The
MOC4 is an extension of the 3rd. release (MOC3) and contains about
twice more observations than its predecessor. However, many of these
new observations were obtained in non photometric conditions%
\footnote{Observations with declination $\left|\delta\right|<1.26^{\circ}$
or galactic latitude $\left|b\right|<15^{\circ}$; see
\url{http://www.astro.washington.edu/users/ivezic/sdssmoc/sdssmoc.html }.}
and were excluded from our analysis. This clean up results in a final
sample of 94,116 observations, corresponding to 70,234 asteroids.

There are different approaches to search for specific taxonomic classes
among a large set of observations like the SDSS-MOC4. \citet{2006Icar..183..411R},
for example, selected candidate V-type asteroids by applying a Principal
Component Analysis to the distribution of the observations. \citet{2010A&A...510A..43C}
used a method based on comparison of the whole set of colors with
templates of the known classes. \citet{2008Icar..198...77M} compared
individual colors to those of known V-type asteroids. \citet{2012Icar..220..577S}
selected V-type candidates studying the distribution of the observations
in a color-color diagram. It is worth noting that these approaches
are all independent and produce results that show a good overlapping.
In principle, it is not possible to say that one method should be
preferred over another.

In this work, we followed the approach of \citet{2006Icar..183..411R}
that basically consists of the following steps: (i) for each observation
in our sample, we computed the photometric reflectance fluxes normalized
to the $r$ band with their corresponding errors; (ii) we discarded
the observations with errors larger than 10 \% in any of the fluxes;
(iii) we applied the Principal Components Analysis to the remaining
observations and chose those with principal components $PC_{1}<0$
and $PC_{2}<-0.158$; and (iv) from this last sample, we selected
the observations linked to asteroids having osculating $1.3<q<1.77$
AU. At this point, it is important to clarify the criteria used to
select the observations. The criterion in step (iii) is more relaxed
than the one applied by \citet{2006Icar..183..411R} to select V-type
candidates, and the result is probably more contaminated by candidates
of other taxonomic classes, especially the S and Q classes that can
be mistaken with the V class in the SDSS photometry. Nevertheless,
our aim is not to make a precise classification of the observations
but to produce a list of potential targets for spectroscopic observations,
and a more restrictive selection criterion could discard targets that
might be interesting. The criterion in step (iv) is also more relaxed
than the actual definition of Mars crossers. However, the choice of
1.77 AU as the upper limit of the population's perihelion distance,
instead of the 1.666 AU classical limit, allows us to include
some IMB asteroids that may get to $q<1.66$ AU in a few hundreds
of thousand years. In fact, most IMB asteroids currently in the interval
$1.67<q<1.77$ AU evolve in a Kozai-like regime, which produces large
amplitude coupled oscillations of the asteroid's eccentricity and
inclination, with periods of the order of $10^{5}$ years, while the
semi-major axis remains fixed (e.g. \citealp{2011epsc.conf..402V}).

Our procedure led to 30 observations matching the above criteria,
which correspond to 18 asteroids listed in Table \ref{table1}. This
table gives the orbital parameters and absolute magnitude of the asteroids,
the values of the principal components and the number of Sloan observations
used. It also lists the $(i-z)$ color index, which gives an idea
of the depth of the absorption band long-wards of 0.75 $\mu$m. Comparing
this list with the taxonomy developed by \citet{2010A&A...510A..43C}%
\footnote{Available at the Planetary Data System Node, 
\url{http://sbn.psi.edu/pds/resource/sdsstax.html }.},
we found 12 asteroids with at least one observation in the SDSS-MOC4
identified by these authors as belonging to the V class, while 4 candidates
were classified by them as Q-type and 2 remain unclassified. We
also found the two asteroids reported by \citet{2010Icar..208..773M}
from NIR observations. This information is summarized in the last
column of Table \ref{table1}.

\subsection{Spectroscopic observations of V-type candidates}
\label{sec2-2}

\begin{table*}
\caption{Observational circumstances of the V-type candidates observed with
the SOAR telescope. JD indicates the time of the beginning of the
observation. Air mass and V apparent magnitude are listed in the second
and third columns. $\phi,$ $R$ and $\Delta$ are the solar phase
angle, the heliocentric distance and the geocentric distance of the
target, respectively.}
\label{table2}
\bigskip{}
\centering{}%
\begin{tabular}{ccccccc}
\hline 
Asteroid number & JD ($2455000+$) & Air mass & $V$ & $\phi$ {[}$^{\circ}${]} & $R$ {[}AU{]} & $\Delta$ {[}AU{]}\tabularnewline
\hline 
(1468) & $929.78819039$ & $1.46$ & $18.20$ & $19.9$ & $2.79$ & $2.36$\tabularnewline
(31415) & $804.49685336$ & $1.16$ & $18.51$ & $28.6$ & $2.09$ & $1.97$\tabularnewline
(32008) & $929.74843229$ & $1.31$ & $17.19$ & $15.0$ & $2.32$ & $1.47$\tabularnewline
(130988) & $804.80536412$ & $1.14$ & $16.84$ & $9.4$ & $1.83$ & $0.84$\tabularnewline
\hline 
\end{tabular}
\end{table*}

\begin{table*}
\caption{Solar analog stars used to obtain the reflectance spectra of the asteroids.
The last column indicates the asteroid's spectrum to which the solar
analog correction was applied. Note that HD 102365 was observed at
different air masses for different asteroids.}
\label{table3}
\bigskip{}
\centering{}%
\begin{tabular}{cccccccc}
\hline 
{\scriptsize{Star name}} & {\scriptsize{$\alpha$ (J2000)}} & {\scriptsize{$\delta$ (J2000)}} & {\scriptsize{Air mass}} & {\scriptsize{$V$}} & {\scriptsize{$B-V$}} & {\scriptsize{Spectral type}} & {\scriptsize{Asteroid(s)}}\tabularnewline
\hline 
{\scriptsize{HD 102365}} & {\scriptsize{$11^{\mathrm{h}}$$46^{\mathrm{m}}$$31.07^{\mathrm{s}}$}} & {\scriptsize{$-40^{\circ}$$30^{\prime}$$01.27^{\prime\prime}$}} & {\scriptsize{1.15 / 1.32}} & {\scriptsize{4.89}} & {\scriptsize{0.66}} & {\scriptsize{G2V}} & {\scriptsize{(1468) / (31415)}}\tabularnewline
{\scriptsize{HR 3538}} & {\scriptsize{$08^{\mathrm{h}}$$54^{\mathrm{m}}$$17.95^{\mathrm{s}}$}} & {\scriptsize{$-05^{\circ}$$26^{\prime}$$04.06^{\prime\prime}$}} & {\scriptsize{1.12}} & {\scriptsize{6.00}} & {\scriptsize{0.67}} & {\scriptsize{G3V}} & {\scriptsize{(32008)}}\tabularnewline
{\scriptsize{HR 8700}} & {\scriptsize{$22^{\mathrm{h}}$$53^{\mathrm{m}}$$37.93^{\mathrm{s}}$}} & {\scriptsize{$-48^{\circ}$$35^{\prime}$$53.83^{\prime\prime}$}} & {\scriptsize{1.21}} & {\scriptsize{6.03}} & {\scriptsize{0.63}} & {\scriptsize{G0V}} & {\scriptsize{(130988)}}\tabularnewline
\hline 
\end{tabular}
\end{table*}

Four asteroids in our list of V-type candidates have been observed
in service mode using the Goodman HT Spectrograph installed at the
4.1 m Southern Astrophysical Research Telescope (SOAR) in Cerro Pachón,
Chile, during semester 2012A. The observational circumstances are
summarized in Table \ref{table2}. The Goodman HT Spectrograph was
used in single long slit mode, with a slit of 1.03'' in width. It
was equipped with a blocking filter GG-385 and a grating of 300 l/mm,
which gives a resolution $R\sim1390$. The nominal wavelength
coverage of the optical system is from 0.39 to 0.91 $\mu$m.
The exposure time per target varies between 550 and 600 sec, which
allowed to get a S/N of up to $\sim30$ in the best case.

To subtract the solar contribution from the spectra, solar analog
stars have been also observed at approximately the same air mass of
the asteroids. The exposure time was of 1 sec for each star,
and the star spectrum was obtained immediately before or after the
asteroid's spectrum. The stars used are listed in Table \ref{table3}.
Bias, dome flat field, and Hg-Ar lamp images have been also obtained
for calibration and reduction purposes. The spectra have also
been corrected by atmospheric extinction, which introduces some unreddening
only for the objects observed at the large air masses, although very
well inside the 1-$\sigma$ dispersion of the spectra.

\begin{figure*}
\centering{}%
\includegraphics[width=0.49\textwidth]{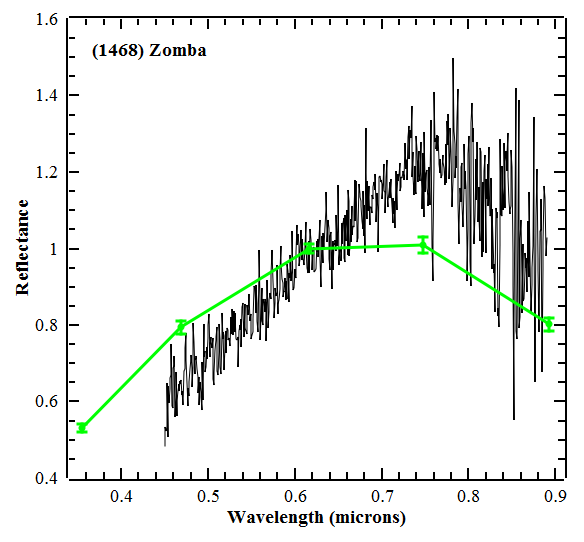}
\includegraphics[width=0.49\textwidth]{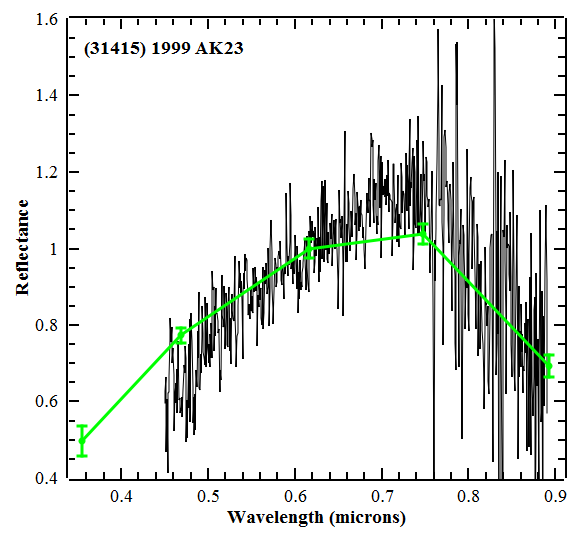}
\includegraphics[width=0.49\textwidth]{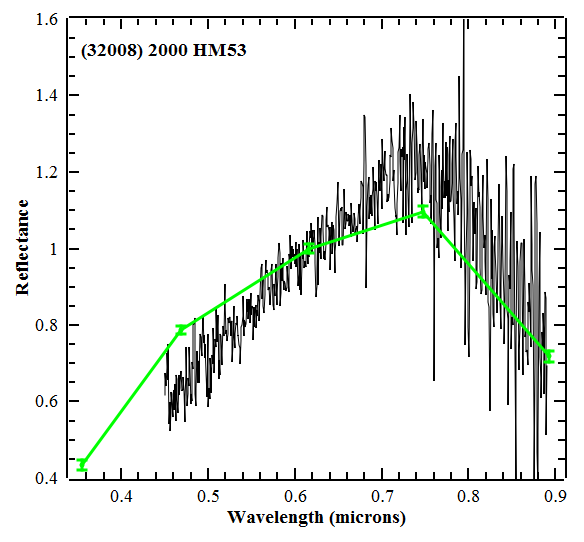}
\includegraphics[width=0.49\textwidth]{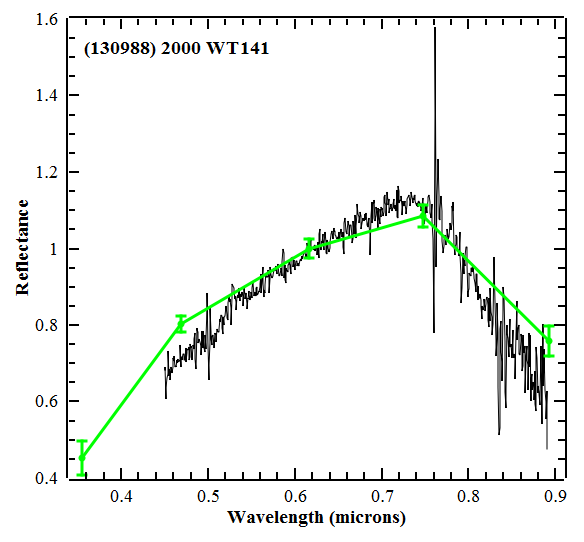}
\caption{Spectra (black lines) of the four asteroids observed with the SOAR
telescope. Green dots represent the SDSS observations with their corresponding
error bars. The reflectance is normalized to 1 at 0.616 $\mu$m.}
\label{figure1}
\end{figure*}

The resulting spectra for each of the four asteroids are shown in
Fig. \ref{figure1}. The spectra have been normalized to reflectance
1 at the center of the $r$ band (0.616 $\mu$m), instead of the classical
value of 0.55 $\mu$m, to make them comparable to the SDSS data. The
spectra have also been limited to the interval from 0.45 to 0.89 $\mu$m,
since the data below 0.45 $\mu$m is quite noisy and it is actually
irrelevant for taxonomic classification purposes (see next paragraph),
and the data above 0.89 $\mu$m is extremely noisy and unusable.
The small dots linked by green lines represent the SDSS fluxes, with
their corresponding error bars. The agreement between the Sloan fluxes
and the spectrum is good, except in the case of (1468) Zomba. 

In order to verify if the observed spectra belong or not to the V
class, we performed a comparison to a template of the class. The template
was synthesized using 37 V-type spectra in the interval from
0.49 to 0.92 $\mu$m that are available in the SMASS II
survey%
\footnote{Actually, the SMASS II spectra are in the interval from 0.43
to 0.92 $\mu$m, but they were cut off at 0.49 $\mu$m to make them
compatible with the S$^{3}$OS$^{2}$ ones.} 
(\citealp{2002Icar..158..106B}, \nocite{2002Icar..158..146B}b) and
the S$^{3}$OS$^{2}$ survey \citep{2004Icar..172..179L}, according
to the following procedure: First, all the spectra were rebinned in
wavelength into 18 channels of 0.025 $\mu$m each. If $f_{i,k}$ denotes
the mean flux of the $i$-th spectrum at the $k$-th channel and $\sigma_{i,k}$
is the corresponding standard deviation, then the template's flux
at the $k$-th channel is computed by the weighted average:
\[
\bar{f}_{k}=\frac{\sum_{i}\: f_{i,k}\sigma_{i,k}^{-2}}{\sum_{i}\:\sigma_{i,k}^{-2}}
\]
where the summation runs over the 37 reference spectra. The corresponding
unbiased weighted sample variance and standard deviation $\bar{\sigma}_{k}$
are also computed. The membership of any spectrum to the template's
class is then determined by computing the weighted distance:
\[
d_{i}=\left[\frac{\sum_{k}\left(f_{i,k}-\bar{f}_{k}\right)^{2}\left(\sigma_{i,k}^{2}+\bar{\sigma}_{k}^{2}\right)^{-1}}{\sum_{k}\left(\sigma_{i,k}^{2}+\bar{\sigma}_{k}^{2}\right)^{-1}}\right]^{1/2}
\]
where the summation runs over the 18 channels. Among the 37 reference
spectra, the maximum distance to the template is $d_{\mathrm{max}}=0.038$,
therefore any spectrum having $d_{i}\leq d_{\mathrm{max}}$ can be
considered to belong to the V class. 

\begin{figure}
\centering{}%
\includegraphics[width=0.8\columnwidth]{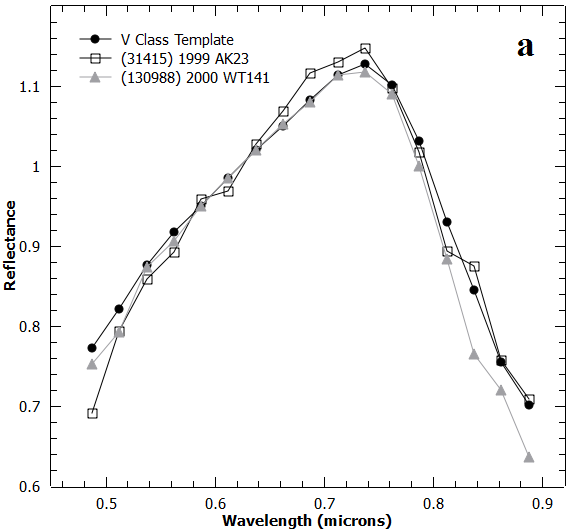}
\includegraphics[width=0.8\columnwidth]{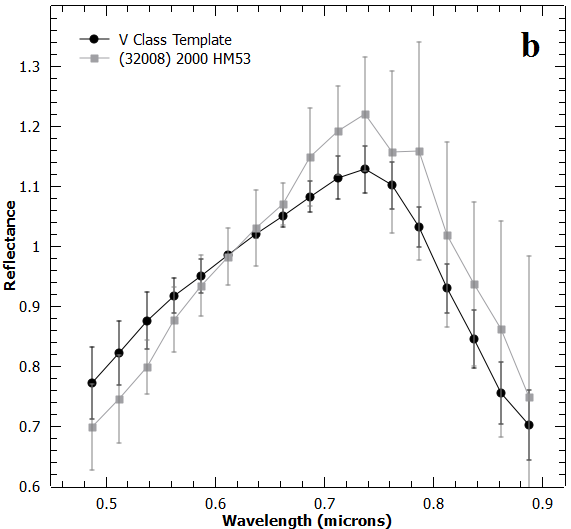}
\includegraphics[width=0.8\columnwidth]{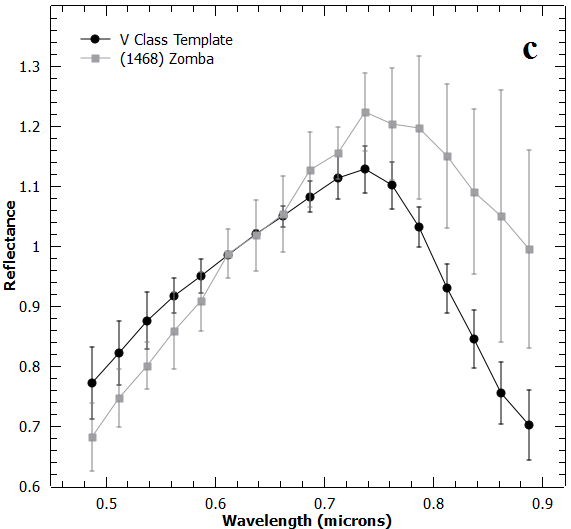}
\caption{Comparisons between the V class template (full black dots) and the
rebinned spectra of the observed asteroids (other symbols). In panels
b and c, we also show the corresponding 1-$\sigma$ error bars of
the spectra and the template. Note the different vertical scale in
panel a.}
\label{figure2}
\end{figure}

Table \ref{table4} gives the distance to the template of the four
spectra observed in this work. In the case of (31415) and (130988),
we conclude that they can be classified as V-type asteroids. Figure
\ref{figure2}a shows a comparison between the two rebinned spectra
and the class template indicating a very good agreement. 

In the case of (32008), the value of $d_{i}$ would indicate that
this in principle may not be a V-type. However, the comparison of
the rebinned spectrum to the class template, shown in Fig. \ref{figure2}b,
indicates a profile compatible with a V-type spectrum, especially
concerning the deep absorption feature long-wards of 0.75 $\mu$m.
We may also note from this comparison that, even if the spectrum is
overestimated with respect to the template beyond the normalization
wavelength, all the channels are indistinguishable within the 1-$\sigma$
uncertainties. Thus, (32008) can be marginally classified as a V-type. 

\begin{table}
\caption{Distance $d_{i}$ between the spectrum of each asteroid and the V
class template. Spectra with $d_{i}\leq0.038$ are undoubtfully classified
as V-type.}
\label{table4}
\bigskip{}
\centering{}%
\begin{tabular}{lc}
\hline 
Asteroid & $d_{i}$\tabularnewline
\hline 
(1468) & 0.080\tabularnewline
(31415) & 0.024\tabularnewline
(32008) & 0.050\tabularnewline
(130988) & 0.015\tabularnewline
\hline 
\end{tabular}
\end{table}

\begin{figure}
\centering{}%
\includegraphics[width=1.0\columnwidth]{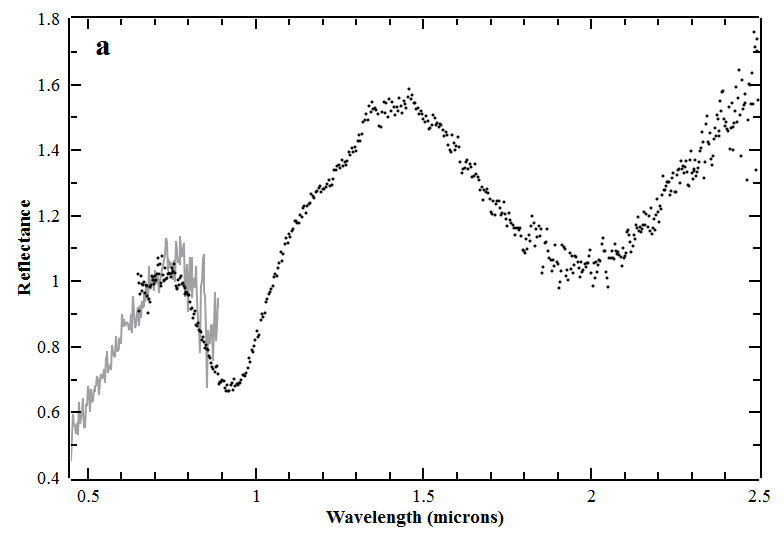}
\includegraphics[width=1.0\columnwidth]{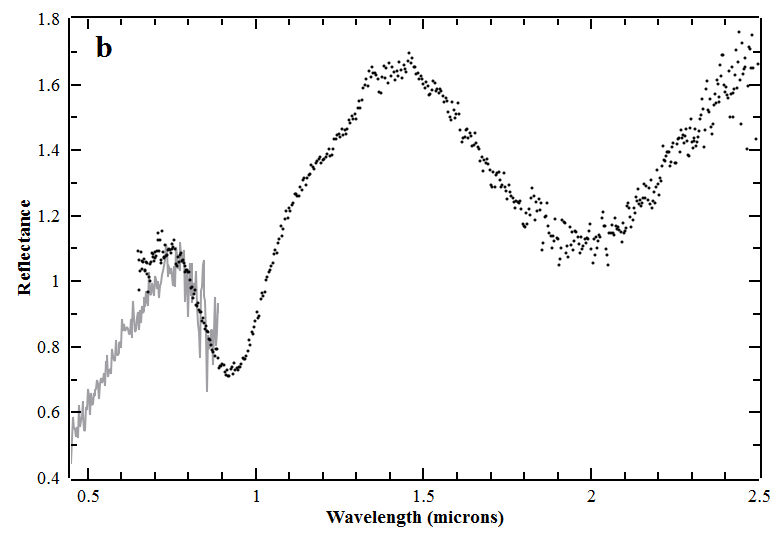}
\caption{Spectrum of (1468) in the range 0.45--2.5 $\mu$m, obtained
by collating our visible spectrum (gray line) with a NIR spectrum of
\citet{2010Icar..208..773M} (black dots). We have applied a 20:1 rebinning
in wavelength to the visible spectrum. The reflectance has been normalized
to 1 at two common wavelengths: 0.7 $\mu$m (panel a) and 0.8 $\mu$m
(panel b). In either case, it is not possible to find a good match
between the visible and NIR spectra.}
\label{figure3-1}
\end{figure}

In the case of (1468), the distance $d_{i}$ is incompatible with
the class, and the rebinned spectrum is distinguishable from the template
within the 1-$\sigma$ uncertainties (Fig. \ref{figure2}c). Therefore,
this asteroid cannot be classified as a V-type through this procedure.
We recall, however, that \citet{2010Icar..208..773M} reported two
observations of (1468) in the NIR%
\footnote{Available at the Planetary Data System Node,
\url{http://sbn.psi.edu/pds/resource/irtfspec.html }.}
These spectra cover the interval 0.65--2.5 $\mu$m and 0.82--2.5
$\mu$m, respectively, and the latter one allows those authors to
perform a mineralogical analysis and a comparison to the HED meteorites,
concluding that (1468) has a basaltic composition. In Fig. \ref{figure3-1},
we show an attempt to join our visible spectrum (rebinned) with this
NIR spectrum, using two different common wavelengths for normalization:
0.7 and 0.8 $\mu$m. Since it is not possible to find a match, we
conclude that our visible spectrum might have been affected by some
instrumental problem producing a significant reddening. Among the
possibilities, a target partially outside the slit during the acquisition
of the spectrum might produce such effect.

Regardless of the issue with (1468), our results provide
the first confirmations of V-type SDSS candidates among the
Mars crossers. In the cases of (31415) and (32008), unfavorable observational
conditions ((31415) was too faint and (32008) was observed at a rather
large air mass) produced very noisy spectra, which implies larger
uncertainties in the taxonomic classification. Therefore, although
these two asteroids can be classified as V-type following our procedure,
better spectroscopic observations would be advisable to reinforce
this classification.

\section{Orbital evolution of V-type Mars crossers}
\label{sec3}

Of the three MC asteroids classified as V-type by our spectroscopic
observations, two ((32008) and (130988)) are not currently Mars crossers
but they are slightly above the limit in perihelion distance used
by the classical definition of this population. The third one, (31415),
is currently a Mars crosser but it is also very close to the upper
limit of the population's $q$. In order to check the orbital behavior
of these bodies, we have followed their evolution as test particles
using the symplectic integrator SWIFT, modified to properly manage
close encounters of the asteroids with the planets (SKEEL code; 
\citealp{2000AJ....120.2117L}).
Our model also includes the effect of Yarkovsky forces on the asteroids
using the same approach as in \citet{2008Icar..193...85N}, i.e. we
introduce a dissipative force which is parallel to the orbital velocity
and produces a prescribed drift $da/dt$ in semi-major axis. The maximum
allowed drift for an asteroid is given by
\[
\left|\frac{da}{dt}\right|=k\frac{1\,\mathrm{km}}{D}
\]
where $k$ is a parameter that for V-type asteroids is approximately
$2.5\times10^{-10}$ AU/y (see \citet{2008Icar..193...85N} for details),
and $D$ is the asteroid's diameter in km. Since the albedo
of these three asteroids is unknown, the diameter in this case
was estimated from the absolute magnitude $H$ assuming an
albedo of 0.35. This is compatible with recent estimates of (4) Vesta's
albedo (\citealp{2011A&A...533L...9F}), and with the mean albedo
of Vesta family members from the WISE/NEOWISE survey (\citealp{2012ApJ...759L...8M}).
The values are summarized in Table \ref{table6}. The Solar System
model considered all the 8 planets from Mercury to Neptune, and each
simulation covered a time span of at least 200 My, but some simulations
were extended up to 1 Gy.

\begin{table}
\caption{Estimated diameters and maximum Yarkovsky drifts.}
\label{table6}
\centering{}%
\begin{tabular}{lcc}
\hline 
Asteroid & $D$ {[}km{]} & $\left|\dot{a}\right|_{\mathrm{max}}$ {[}AU/y{]}\tabularnewline
\hline 
(31415) & 2.96 & $8.5\times10^{-11}$\tabularnewline
(32008) & 3.40 & $7.4\times10^{-11}$\tabularnewline
(130988) & 1.96 & $12.8\times10^{-11}$\tabularnewline
\hline 
\end{tabular}
\end{table}

\begin{figure}
\centering{}%
\includegraphics[width=0.75\columnwidth]{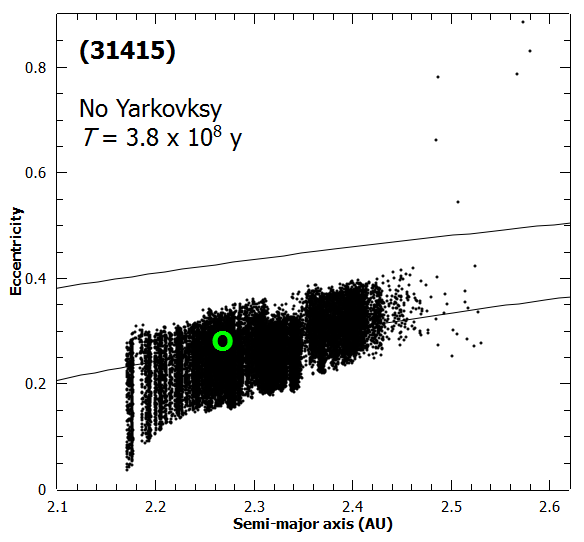}
\includegraphics[width=0.75\columnwidth]{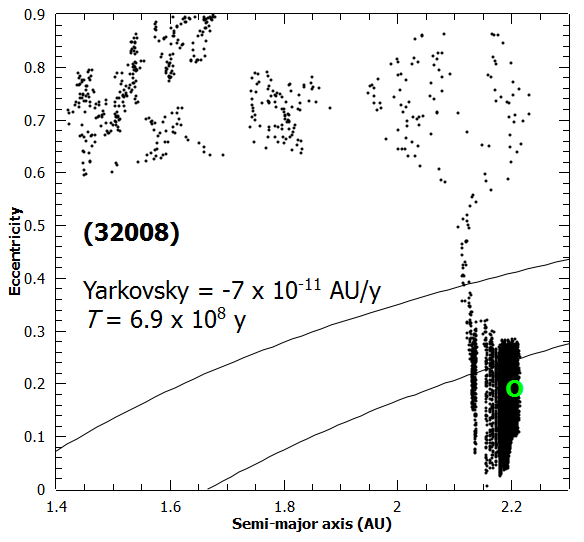}
\includegraphics[width=0.75\columnwidth]{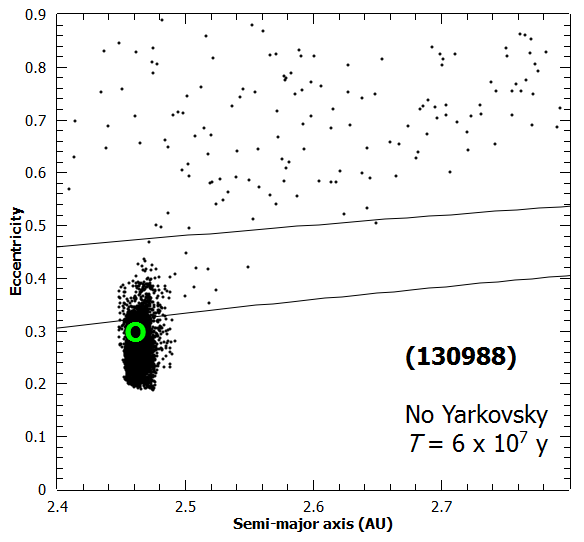}
\caption{Examples of the orbital evolution (black dots) of the three V-type asteroids
confirmed in this study, in the space of semi-major axis vs. eccentricity.
All the examples shown here led to Earth crossing orbits. The corresponding
drift in semi-major axis due to the Yarkovsky effect is indicated,
when applicable. The time $T$ is the approximate time when the orbit
enters the Mars crossing region to be quickly driven to an Earth crossing
orbit. The full black lines represent the limits of the MC region. The
area above the uppermost line corresponds to the EC region. The grey
open circle indicates the current location of the asteroids, which was used
as initial condition for the simulations.}
\label{figure3}
\end{figure}

For each of the three asteroids, we performed 11 simulations starting
with the same orbital initial conditions: 1 simulation with no Yarkovsky
effect, 5 with a Yarkovsky effect causing positive drifts in semi-major
axis (randomly sorted within the maximum allowed drift rate), and
5 with a Yarkovsky effect causing negative drifts. All the simulations
show that the asteroids temporarily acquire $q<1.666$ AU several
times during their evolution, jumping in and out of the MCs region.
The orbits of (31415) and (32008) appear to be quite stable and at
most 18 \% of the simulations led to paths into the EC region only
after several 100 My of evolution, with the subsequent elimination
of the asteroid due to close encounters with the Earth or Venus. Examples
of such simulations are shown in Fig. \ref{figure3}. In these cases,
the transition to the Earth crossing regime is very fast, lasting
of the order of 10 My or less.

On the other hand, in the case of asteroid (130988), we found
that 72 \% of the performed simulations led to a behavior that drives
the asteroid to an Earth crossing orbit, eventually discarding it
after a close approach to the Earth. The delivery of this asteroid
to the EC region may happen as early as after 60 My of evolution.
An example of this behavior is also shown in Fig. \ref{figure3}.
Asteroid (130988) would be the first confirmed case of a V-type asteroid
that originates in the IMB, passes through the MC region and arrives
to the near-Earth space. Again in these cases, the evolution in the
Mars crossing and Earth crossing regimes lasts $\sim10$ My or less. 

\section{Conclusions}
\label{sec4}

In this work, we applied the method of \citet{2006Icar..183..411R}
to identify candidate V-type asteroids among the populations of Mars
crossers, using the photometry of the SDSS Moving Objects Catalog
(4th. release). We identify 18 asteroids with colors compatible with
those of the V class, and performed spectroscopic observations of
4 of these candidates during semester 2012A with the SOAR Telescope.
The resulting spectra indicate that:
\begin{itemize}
\item two candidates are undoubtfully classified as V-type asteroids; 
\item one candidate can be marginally classified as a V-type;
\item the fourth candidate cannot be classified as a V-type according
to our observations, although according to \citet{2010Icar..208..773M}
it has a basaltic composition;
\item one of the confirmed V-type asteroids fully enters the Mars crossing
region after some tens of million years of evolution, and is quickly
driven (in less than 10 My) to have a close encounter with the Earth;
\item the other two confirmed V-type asteroids temporarily enter the Mars
crossing region several times during their orbital evolution, but
they have quite stable orbits and are hard to be driven to the Earth
crossing region over hundreds of million years of evolution.
\end{itemize}
Our observations provide the first spectroscopic confirmation 
of V-type SDSS candidates among the Mars crossers. They
help to enlarge the database of known basaltic asteroids among this
population, and give support to the Mars crossing regime as an alternative
path to deliver basaltic meteorites to the Earth. However, the actual
efficiency of this dynamical path, and the still much lesser amount
of V-type asteroids in the MC population relative to the EC population
still remain open questions. 

\section*{Acknowledgments }

The authors wish to thank V. Reddy and an anonymous reviewer
for their helpful comments and suggestions. FR also wish to thank
G. Valsecchi and G. Gronchi for the discussion about a good dynamical
the criterion to define Mars crossers. This work has been supported
by Coordenação de Aperfeiçoamento para Pessoal de Nível Superior (CAPES,
Brazil), Conselho Nacional de Desenvolvimento Científico e Tecnológico
(CNPq, Brazil), and Consejo Nacional de Investigaciones Científicas
y Técnicas (CONICET, Argentina).

\end{document}